\documentclass[11pt]{article}
\usepackage{amsmath}
\usepackage{graphicx}
\usepackage{amssymb}
\usepackage{amsfonts}
\usepackage{latexsym}

\def\bea{\begin{equation}}
\def\eea{\end{equation}}
\def\beq{\begin{eqnarray}}
\def\eeq{\end{eqnarray}}

\def\ln{\,\mbox{ln}\,}
\def\Det{\,\mbox{Det}\,}

\def\tr{\,\mbox{tr}\,}
\def\Tr{\,\mbox{Tr}\,}

\def\Box{\square}

\def\al{\alpha}
\def\be{\beta}

\def\ga{\gamma}

\def\ep{\epsilon}

\def\na{\nabla}

\def\om{\omega}

\def\Ga{\Gamma}

\begin{document}

\title{Form factors and non-local Multiplicative Anomaly
for fermions with background torsion}
\author{G. de Berredo-Peixoto\footnote{Email address:
guilherme@fisica.ufjf.br}$\;$ and $\,$
A. Espinosa Maic\'a\footnote{Email address: maica@fisica.ufjf.br} \\ \\
Departamento de F\'{\i}sica, ICE, Universidade Federal de Juiz de Fora
\\
Campus Universit\'{a}rio - Juiz de Fora, MG, Brazil  36036-330}

\date{}
\maketitle

\begin{quotation}
\noindent
\begin{center}
{\large{\bf Abstract}} \\
\end{center}

We analyse the Multiplicative Anomaly (MA) in the case of
quantized massive fermions coupled to a background
torsion. The one-loop Effective Action (EA) can be expressed in terms
of the logarithm of determinant of the appropriate first-order
differential operator acting in the spinors space. Simple
algebraic manipulations on determinants must be used in order
to apply properly the Schwinger-DeWitt technique, or even the
covariant perturbation theory (Barvinsky and Vilkovisky,
1990), which is used in the present work. By this method, we
calculate the finite non-local quantum corrections, and
analyse explicitly the breakdown of those algebraic
manipulations on determinants, called by MA. This feature
comes from the finite non-local EA, but does not affect
the results in the UV limit, in particular the beta-functions.
Similar results was also obtained in previous papers but for
different external fields (QED and scalar field).


\end{quotation}
%\newpage

%%%%%%%%%%%%%%%%%%%%%%%%%%%%%%%%%%%%%%%%%%%%%%%%%%%%%%%%%%%%%%%%%
\section{Introduction}

The Effective Action (EA) can be considered as an important tool for
studying quantum effects in field theories.
It should be derived for a given Quantum Field Theory (QFT)
taking into account its ambiguities: one has to distinguish
unambiguous physical effects from the technical features in the
calculations.

A relevant ambiguity in QFT concerns, for example, the dependence
on the renormalization point, which is described by the parameter
$\mu$ in the Minimal Subtraction (MS) scheme adopted in the renormalization
procedure. Another ambiguity is the gauge-fixing dependence, which, in
principle, can be eliminated on-shell. One can say that many properties
obtained in studying the running couplings parameters depend on
the renormalization scheme. For example, the renormalization
group $\be$-functions in massive theories calculated in MS scheme
are different from the ones calculated using the more physical momentum
subtraction scheme. In the low energies regime, the approach of momentum
subtraction enables one to observe the decoupling phenomenon, which in QED
is known as the Appelquist and Carazzone theorem \cite{AC}.

There are other ambiguities in quantum contributions, but we can
mention an interesting property of the UV divergences: the leading
logarithmic divergences define the most stable and universal part
of quantum corrections. Let us point out that the UV limit of
$\be$-functions, which is strongly related to UV-divergences,
does not depend on the renormalization scheme. On the other side,
the calculational techniques, valid for studying the unambiguous
UV part (specifically the logarithmic divergences), fails when
one investigates the finite non-leading part of one-loop EA,
which present then some ambiguity. This kind of ambiguity, coming
from what is called the non-local {\it multiplicative anomaly} (MA), was
treated in recent papers, where different examples was studied:
(i) finite 1-loop quantum corrections from massive fermionic fields
in electromagnetic background in curved space, Ref. \cite{bruno};
and (ii) finite 1-loop quantum corrections from massive fermionic fields
in Yukawa model (also in curved space), Ref. \cite{dante}. It is worth mentioning
that this non-local MA is different from what is investigated by many authors in
previous papers \cite{MA-1,MA-2,MA-3,MA-4}, by using the
$\zeta$-regularization. In those works, the MA can be related to
the ambiguity of the choice of $\mu$ itself \cite{MA-3,MA-4}. One can
say that the MA which we are considering is a non-local version
of earlier MA. In this sense, we are dealing with a different kind
of ambiguity of EA.

In this paper, we are going to investigate the MA in the
finite 1-loop quantum corrections of massive Dirac fields
in the background of an axial vector field, along with gravity.
The relevance of this issue is obvious by the great number
of research papers on CPT/Lorentz violating theories \cite{collakost}
and also on theories with torsion \cite{hehl}. In the minimal
coupling between fermions and torsion, only the axial component
of torsion takes position in the mathematical description
(see the recent review \cite{reviewsh}), thus an axial background
field is useful not only to describe CPT/Lorentz violation, but also
to describe torsion issues, ranging from atomic physics to aspects
on Quantum Field Theory, Cosmology and Astrophysics. It does not
matter if one considers torsion effects on Cosmology or CPT/Lorentz
violating theories: in most interesting cases, quantum corrections
of quantized matter play relevant role in physical
phenomena\footnote{This scenario is not sensitive to the fact
that torsion quantization meets serious problems of
renormalizability and unitarity \cite{2000}.}, such
that the MA in finite 1-loop EA can not be thought as a mathematical
feature without physical consequences.

This paper is organized as follows. In Section 2, we describe briefly
the relevant aspects of the covariant perturbation theory which are
useful for calculating the non-local MA. This calculation is performed
in Section 3, for massive fermions coupled to an external axial vector.
The results are given explicitly and in Section 4 we deserve some attention
to the effects of MA in the beta functions. Finally, in Section 5, we
draw our final comments and conclusions.

\section{Preliminary considerations}

Let us consider the quantum corrections coming from free massive
fermion fields coupled to the background torsion\footnote{We mean that they are
coupled to an external axial vector.}. The action reads
$$
S_f = \int d^4x \sqrt{-g}\,\,{\bar \psi} \hat{H} \psi\,,
$$
where
\beq
\hat{H} = i\ga^{\mu}\na_{\mu} + m\hat{1} + \eta\ga^5\ga^\mu S_\mu\,.
\label{act}
\eeq
Here $\hat{1}$ is the identity operator,
$S_\mu$ is the axial vector, which can be identified with the $b_\mu$
parameter in CPT/Lorentz violating theories \cite{collakost} and also
with external torsion, and $\eta$ is the non-minimal coupling parameter
(the value $\eta = 1/8$ describes minimal coupling with torsion).

The one-loop EA can be understood as the classical
action of background fields plus the one-loop correction (the subject
of our interest) \cite{book}:
\beq
{\bar \Ga}^{(1)} \,=\,-\, \ln \Det \hat{H}\,.
\label{Det}
\eeq

Here and below, we use Euclidean signature.
We shall compute the expression (\ref{Det}) through the heat-kernel
method and the Schwinger-DeWitt technique (see Refs. \cite{book,dewitt,avramidi}),
and this requires reducing the problem to the derivation of
$\,\ln \Det\,\hat{\cal O}$,
where the general form for $\hat{\cal O}$ can be found as
\beq
\hat{{\cal O}} \,=\,
{\widehat{\Box}} + 2{\hat h}^\mu\na_\mu + {\hat \Pi}\,.
\label{O}
\eeq
This method of calculation is not directly suited for the first
order differential operator $\hat{H}$. However, one can overcome
this difficulty by multiplying $\hat{H}$ by another operator $\hat{H}_1$
and using the relation
\beq
\ln \Det\,\hat{H} \,=\,\ln \Det\,(\hat{H}\,\hat{H}_1)
\,-\,\ln \Det\,\hat{H}_1\,.
\label{rel}
\eeq

The most obvious choice is of course\footnote{For practical
purposes and calculational convenience, this is realized by considering
$\hat{H}_1 = -\hat{H}$. The difference will be just a constant
term which does not change the main results.} $\hat{H}_1 = \hat{H}$,
so that $\ln \Det\,\hat{H}$ is obtained directly by calculation of
$(1/2) \ln \Det\,(\hat{H}.\hat{H})$. It turns out, however, that
this procedure makes computation much more complicated than the
one made by an appropriate choice for $\hat{H}_1$, say,
\beq
\hat{H_1} = -i\ga^{\mu}\na_{\mu} + m\hat{1} + \eta\ga^5\ga^\mu S_\mu\,.
\label{H1}
\eeq
It is possible to consider also different possibilities, such as
\beq
\hat{H_2} = -i\ga^{\mu}\na_{\mu} + m\hat{1}\,,
\label{H2}
\eeq
\beq
\hat{H_3} = -i\ga^{\mu}\na_{\mu} - m\hat{1}\,.
\label{H3}
\eeq

The one-loop EA is related to the coincidence limits
\beq
\lim\limits_{x \to x^\prime}{\hat a}_k(x,x^\prime)
\,=\,{\hat a}_k\big| \,,
\label{SchDW-6}
\eeq
which appear in its proper-time expansion (see more details
in Ref. \cite{dewitt,avramidi}) throughout
\beq
\bar{\Gamma}^{(1)} = -\frac{1}{2} \Tr\,\int\limits_0^\infty
\frac{ds}{s}\hat{U}_0(x,\,x^\prime ;\,s)\sum_{k=0}^{\infty}
(is)^k\hat{a}_k(x,\,x^\prime)\,,
\label{EAa}
\eeq
where $\Tr$ means integration in $x$ of quantities in the limit
$x\to x^\prime$ as well as the ordinary trace in discrete labels
(i.e., $\Tr = \int d^4x \lim\limits_{x\to x^\prime}\tr$),
the minus sign is present because of the Grassmann parity
(we consider here only the fermionic sector), and
$\hat{U}_0(x,\,x^\prime ;\,s)$ is an operator function
related to the geodesic distance between $x$ and $x^\prime$
and the Van-Vleck-Morette determinant.

The coefficients ${\hat a}_k\big|$ are useful to
describe the EA in several field theory models.
For instance, in the two-dimensional spacetime, ${\hat a}_1\big|$
defines logarithmic divergences. In four dimensions, ${\hat a}_1\big|$
describes quadratic divergences, while the logarithmic ones are given by
${\hat a}_2\big|$ (the same happens with ${\hat a}_3\big|$ in six dimensions,
and so on so forth). In four dimensions, all summation with $k>2$
represent finite contributions in the EA.

Fortunately, we do not need to calculate all the coefficients
to investigate the MA from finite contributions, because
Barvinsky and Vilkovisky \cite{bavi90} and also Avramidi
\cite{avramidi89} obtained an equivalent EA but expressed
as a different summation, in powers of curvature (i.e.,
in powers of quantities with dimension of curvature).
Each coefficient of this new expansion (called form factors)
has all non-localities coming from the finite part.
In Ref. \cite{gorsha}, this method was used to calculate
the complete form factors and $\beta$-functions for some
massive fields at 1-loop level (See also Refs. \cite{milton,guszel}
for similar calculations). The approach developed in Refs.
\cite{bavi90,avramidi89} is sometimes called covariant
perturbation theory.

The Schwinger-DeWitt algorithm enables one to obtain ${\hat a}_1\big|$
and ${\hat a}_2\big|$ for the differential o\-pe\-ra\-tor in the general
form (\ref{O}):
\beq
{\hat a}_1\big| = \hat{P} = \hat{\Pi} + \frac{\hat{1}}{6} R -
\na_\mu \hat{h}^\mu - \hat{h}_\mu \hat{h}^\mu \,,
\eeq
\beq
{\hat a}_2\big| = \frac{\hat{1}}{180}\left(
R_{\mu\nu\al\be}^2 - R_{\mu\nu}^2 + \Box R \right)
+\frac{1}{2} \hat{P}^2 + \frac{1}{6} (\Box \hat{P})
+ \frac{1}{12} \hat{S}_{\mu\nu}^2\,,
\eeq
where we denote $R_{\mu\nu\al\be}^2 = R_{\mu\nu\al\be}R^{\mu\nu\al\be}$ (and so on)
and
$$
\hat{S}_{\mu\nu} = \hat{1}[\,\na_\mu\,,\,\na_\nu\,] +
\na_\mu \hat{h}_\nu - \na_\nu \hat{h}_\mu +
\hat{h}_\mu \hat{h}_\nu - \hat{h}_\nu \hat{h}_\mu\,.
$$

\section{Calculation of the one-loop form factors}

Using the technique of covariant perturbation theory, one
can express the one-loop EA (\ref{EAa}) in the form of an
expansion in powers of fields strengths (or curvatures).
Up to second order in curvatures, it can be written as
\beq
\bar{\Gamma}^{(1)} & = & -\frac{1}{2}\int\limits_0^\infty
\frac{ds}{s}
\frac{\mu^{4 - 2\om}}{(4\pi s)^\om}\int d^{2\om}x\sqrt{g}
e^{-sm^2}\tr \left\{ \hat{1} + s\hat{P} +
s^2\left[ R_{\mu\nu}f_1(\tau)R^{\mu\nu} \frac{}{} \right. \right. \nonumber \\
&  & + \left. \left.
Rf_2(\tau)R + \hat{P}f_3(\tau)R + \hat{P}f_4(\tau)\hat{P} +
\hat{S}_{\mu\nu}f_5(\tau)\hat{S}^{\mu\nu}\right]\, \right\}\,,
\label{EA}
\eeq
where $\om$ is the dimensional regularization parameter, $\mu$
is the renormalization parameter with dimension of mass and the
functions $f_i$ of $\tau = -s\Box$ are given by
\beq
f_1(\tau) & = & \frac{f(\tau) - 1 + \tau/6}{\tau^2}\,, \nonumber \\
f_2(\tau) & = & \frac{f(\tau)}{288} + \frac{f(\tau) - 1}{24\tau} -
\frac{f(\tau) - 1 + \tau/6}{8\tau^2}\,, \nonumber \\
f_3(\tau) & = & \frac{f(\tau)}{12} + \frac{f(\tau) - 1}{2\tau}\,,  \\
f_4(\tau) & = & \frac{f(\tau)}{2}  \,, \nonumber \\
f_5(\tau) & = & \frac{1 - f(\tau)}{2\tau}\,, \nonumber
\eeq
with
$$
f(\tau) = \int\limits_0^1 d\al e^{-\al (1 - \al)\tau}\,.
$$

In order to deal with the integrations in parameter $s$, we indicate the works
in Ref. \cite{gorsha}, where the reader can find more technical details. Let us
consider only the terms containing the $S_\mu$ field, thus we disregard the
vacuum contributions (for example, $Rf_2(\tau)R$) as well as the total derivative
terms. Then, we keep attention only to form factors containing $f_3$, $f_4$ and
$f_5$. After cumbersome calculation, one can perform some tricky integrations and
rearrange the relevant terms of (\ref{EA}) to get the following result, for
Euclidean EA in 4 dimensions:
\beq
\bar{\Gamma}^{(1)} & = & -\frac{1}{2(4\pi)^2}\int d^4x \sqrt{g}
\tr\left\{ \hat{P}\left(A + \frac{1}{2\ep} \right) \hat{P} +
\hat{P}\left( (a^2 - 4)\frac{A}{6a^2} - \frac{1}{18}\right) R \right.
\nonumber \\
& & +  \left. \hat{S}_{\mu\nu}\left( \frac{1}{18} + \frac{2A}{3a^2} + \frac{1}{12\ep}\right)
\hat{S}^{\mu\nu}\right\}\,,
\label{EAff}
\eeq
where
$$
A = 1 + \frac{1}{a}\ln \left(\frac{2-a}{2+a}\right)\,; \;\;\;
a^2 = \frac{4u}{u+4}\,; \;\;\; u = -\Box/m^2\,,
$$
and the small parameter $\ep$ was arbitrarily chosen in terms
of $\om$ according to
$$
-\frac{1}{\ep} = \frac{1}{\om - 2} + \gamma +
\ln\left(\frac{m^2}{4\pi\mu^2}\right)\,,
$$
with $\gamma$ denoting the Euler constant. Notice that the coefficient
of $\tr (\hat{P}R)$ has no divergent part ($\propto 1/\ep$).

\subsection{Universality of 1-loop logarithm divergences}

We shall indicate that the MA does not affect the 1-loop
divergences. In order to do so, let us compare, for example,
the two quantities
$$
\Tr\ln(\hat{H}\,\hat{H})|_{{\rm div}}\;\;\;\;
{\rm and} \;\;\;\;
\Tr\ln(\hat{H}\,\hat{H}_1)|_{{\rm div}}\,,
$$
where $\hat{H}$ and $\hat{H}_1$ are given by
Eqs. (\ref{act}) and (\ref{H1}), respectively. Disregarding
the vacuum and superficial contributions, one can calculate
$\Tr\hat{a}_2$, following the Schwinger-DeWitt technique, and
obtain, in $n$ dimensions,
\beq
(\Tr\hat{a}_2 |)_{\hat{H}\hat{H}} & = &
\int d^n x\sqrt{g}\,2^{[n/2]}\,\left\{
\frac{n-2}{12}\eta^2 S_{\mu\nu}^2 - (n-2)(n-3)\eta^2 m^2 S^2
\right. \\
& & \left.
+ \frac{n-4}{6}\left( \eta^2(\na_\mu S^\mu)^2
-\eta^2 R_{\mu\nu}S^\mu S^\nu +
\frac{1}{2}\eta^2 RS^2  + (n-2)\eta^4 S^4\right) \right\}\,,
\nonumber \\
(\Tr\hat{a}_2 |)_{\hat{H}\hat{H}_1} & = &
\int d^n x\sqrt{g}\,2^{[n/2]}\,\left\{ -2\eta^2 m^2 S^2 +
\frac{1}{6} \eta^2 S_{\mu\nu}^2\right\}\,,
\eeq
where $[n/2]$ means the integer part of $n/2$, $S_{\mu\nu}^2 =
S_{\mu\nu}S^{\mu\nu}$, $S^2 = S_\mu S^\mu$ and $S^4 = (S_\mu S^\mu)^2$.
The above two expressions coincide only in four dimensions (when
$\hat{a}_2$ describes logarithm divergences). The same has to
happen with $\hat{a}_1$ in two dimensions, corresponding to
logarithm divergences. Indeed, this is so:
\beq
(\Tr\hat{a}_1 |)_{\hat{H}\hat{H}} & = & \int d^n x\sqrt{g}\,2^{[n/2]}\,
(2-n)\eta^2 S^2 + {\rm vacuum}\,, \\
(\Tr\hat{a}_1 |)_{\hat{H}\hat{H}_1} & = & {\rm vacuum}\,.
\eeq

One could verify this mechanism also for $\Tr\hat{a}_3 |$, but
this issue has already been discussed for other fields
in previous papers \cite{bruno,dante}.

\subsection{Multiplicative Anomaly in four dimensions}

Now we proceed with the calculation of form factors from
Eq. (\ref{EAff}) for four different operators:
$\hat{H}\,\hat{H}$, $\hat{H}\,\hat{H_1}$, $\hat{H}\,\hat{H_2}$
and $\hat{H}\,\hat{H_3}$.
We consider the four dimensional case and omit the
the total derivatives and vacuum terms. By straightforward
computation, we achieve
\beq
-\frac{1}{2} \Tr\ln (\hat{H}\,\hat{H}) & = & -\frac{1}{2(4\pi)^2}
\int d^4x \sqrt{g} \left\{
\eta^2(\na_\mu S^\mu)\left[ k_{\na S}(a) \right]\,(\na_\nu S^\nu) +
\right. \nonumber \\
& & \left. +
\eta^2 S_{\mu\nu}\left[k_{SS}(a) + \frac{2}{3\ep}\right]\,S^{\mu\nu} +
\eta^2 m^2 S_\mu\left[k_S(a) - \frac{8}{\ep} \right]\,S^\mu + \right. \nonumber \\
& & \left. +
\eta^4 S^2\left[ k_{S4}(a) \right]\,S^2\right\}\,,
\label{HH}
\\
-\frac{1}{2} \Tr\ln (\hat{H}\,\hat{H_1}) & = & -\frac{1}{2(4\pi)^2}
\int d^4x \sqrt{g} \left\{
\eta^2 S_{\mu\nu}\left[k_{SS}^{(1)}(a) + \frac{2}{3\ep}\right]\,S^{\mu\nu}
\right. \label{HH1} \\
& & \left. + \eta^2m^2 S_\mu\left[k_S^{(1)}(a)  - \frac{8}{\ep}\right]\,S^\mu
\right\}\,, \nonumber \\
- \Tr\ln (\hat{H}\,\hat{H_2}) & = & -\frac{1}{2(4\pi)^2}
\int d^4x \sqrt{g} \left\{
\eta^2(\na_\mu S^\mu)\left[k_{\na S}^{(2)}(a)\right]\,(\na_\nu S^\nu)
\right.
\nonumber \\
& & \left.
+ \eta^2 S_{\mu\nu}\left[k_{SS}^{(2)}(a) + \frac{2}{3\ep}\right]\,S^{\mu\nu} +
\eta^4 S^2\left[ k_{S4}^{(2)}(a) \right]\,S^2 \right. \nonumber \\
& & \left. + \eta^2m^2 S_\mu\left[k_S^{(2)}(a)  - \frac{8}{\ep}\right]\,S^\mu
\right\} \label{HH2} \\
- \Tr\ln (\hat{H}\,\hat{H_3}) & = & -\frac{1}{2(4\pi)^2}
\int d^4x \sqrt{g} \left\{
\eta^2(\na_\mu S^\mu)\left[k_{\na S}^{(3)}(a)\right]\,(\na_\nu S^\nu)
\right.
\nonumber \\
& & \left.
+ \eta^2 S_{\mu\nu}\left[k_{SS}^{(3)}(a) + \frac{2}{3\ep}\right]\,S^{\mu\nu} +
\eta^4 S^2\left[ k_{S4}^{(3)}(a) \right]\,S^2 \right. \nonumber \\
& & \left. + \eta^2m^2 S_\mu\left[k_S^{(3)}(a)  - \frac{8}{\ep}\right]\,S^\mu
\right\}\,, \label{HH3}
\eeq
where the form factors are, for the first scheme,
\beq
k_{\na S}(a) & = & \frac{16A}{a^2} - 4A + \frac{4}{3} \\
k_{SS}(a) & = & \frac{16A}{3a^2} + \frac{4}{9} \\
k_S(a) & = & \frac{384A}{a^2} - 112A + 32 \\
k_{S4}(a) & = & 16A - \frac{64A}{a^2} - \frac{16}{3}\,,
\eeq
for the second scheme,
\beq
k_{SS}^{(1)}(a) & = &  2A - \frac{8A}{3a^2} - \frac{2}{9} \\
k_S^{(1)}(a) & = & -16A\,,
\eeq
for the third scheme,
\beq
k_{\na S}^{(2)}(a) & = & \frac{8A}{a^2} - 2A + \frac{2}{3} \\
k_{SS}^{(2)}(a) & = &   A + \frac{4A}{3a^2} + \frac{1}{9}\\
k_{S4}^{(2)}(a) & = &  2A - \frac{8A}{a^2} - \frac{2}{3} \\
k_S^{(2)}(a)& = & k_S^{(1)}(a)\,,
\eeq
and finally for the fourth scheme,
\beq
k_{\na S}^{(3)}(a) & = & k_{\na S}^{(2)}(a) \\
k_{SS}^{(3)}(a) & = & k_{SS}^{(2)}(a)  \\
k_{S4}^{(3)}(a) & = & k_{S4}^{(2)}(a) \\
k_S^{(3)}(a) & = & \frac{160A}{a^2} - 56A + \frac{40}{3}\,.
\eeq

By Eq. (\ref{HH}), we can obtain the expression for
$-\Tr\ln (\hat{H}) = -(1/2) \Tr\ln (\hat{H}\,\hat{H})$: it
coincides thus with the expression in the right hand side of
Eq. (\ref{HH}). Now that we have
\beq
\Tr\ln (\hat{H}) = \frac{1}{2}\Tr\ln (\hat{H}\,\hat{H})\,,
\label{H}
\eeq
we obtain easily $\Tr\ln (\hat{H}_1)$ by doing $m\to-m$ and
$S_\mu\to - S_\mu$ in (\ref{H}), what leaves the result unchanged,
of course (it has only even powers on $m$ and $S_\mu$). Thus,
$\Tr\ln (\hat{H}_1) = \Tr\ln (\hat{H})$.

Nevertheless, one can obtain $\Tr\ln (\hat{H_1})$ by substituting the
result (\ref{HH1}) into
\beq
\Tr\ln (\hat{H}_1) = \Tr\ln (\hat{H}\,\hat{H_1}) - \Tr\ln (\hat{H})\,,
\label{trlnh1}
\eeq
what should give the exact expression for $\Tr\ln (\hat{H})$, but instead it
gives
\beq
-\Tr\ln (\hat{H}_1) & = & -\frac{1}{2(4\pi)^2}
\int d^4x \sqrt{g} \left\{
\eta^2(\na_\mu S^\mu)\left[ k_{\na S}^{H_1}(a) \right]\,(\na_\nu S^\nu) +
\right. \nonumber \\
& & \left. +
\eta^2 S_{\mu\nu}\left[k_{SS}^{H_1}(a) + \frac{2}{3\ep}\right]\,S^{\mu\nu} +
\eta^2 m^2 S_\mu\left[k_S^{H_1}(a) - \frac{8}{\ep} \right]\,S^\mu + \right. \nonumber \\
& & \left. +
\eta^4 S^2\left[k_{S4}^{H_1}(a) \right]\,S^2\right\}\,,
\label{finalH1}
\eeq
with
$$
k_{\na S}^{H_1}(a) = - k_{\na S}(a)\,; \;\;\;\;
k_{S4}^{H_1}(a) = - k_{S4}(a)\,,
$$
\beq
k_{SS}^{H_1}(a) & = & - \frac{32A}{3a^2} + 4A - \frac{8}{9} \,, \nonumber \\
k_{S}^{H_1}(a) & = & - \frac{384A}{a^2} + 80A - 32 \,. \nonumber
\eeq
The last result is a definitive indication that relation (\ref{trlnh1})
does not hold, and this is precisely what we call MA. One can ask
of course if relation $\Tr\ln (\hat{H}) = (1/2) \Tr\ln (\hat{H}\,\hat{H})$
itself holds: to detect MA, we must assume it holds, but the contrary hypothesis
would mean that MA is present anyway. The analogous feature in calculating
$\Tr\ln(\hat{H}_2)$ and $\Tr\ln(\hat{H}_3)$ can also be shown as follows.

Let us notice that just by visual inspection, the expressions for
$\Tr\ln (\hat{H}_2)$ and $\Tr\ln (\hat{H}_3)$ can be obtained from
$\Tr\ln(\hat{H})$ by the procedure $S_\mu \to 0$. Then,
\beq
\Tr\ln (\hat{H}_2) = \Tr\ln (\hat{H}_3) =  {\rm vacuum\;\; and\;\; total\;\; derivatives}\,,
\label{trlnh3}
\eeq
so the expressions for $\Tr\ln (\hat{H}\,\hat{H}_{2,3})$ should be both equal to
$\Tr\ln (\hat{H})$ in (\ref{H}) and (\ref{HH}), but they are different as 
shown by relations (\ref{HH2}) and (\ref{HH3}).

To summarize what we have so far: the main object of interest is
$\Tr\ln (\hat{H})$, which can be calculated by four methods,
producing different results, all affected by MA. In the first
method, we used $\Tr\ln (\hat{H}) = (1/2)\Tr\ln (\hat{H}\,\hat{H})$
and got (\ref{HH}). In the second method, we used (\ref{trlnh1}) and
got (\ref{finalH1}). Finally, in the third and fourth methods we
got respectively expressions (\ref{HH2}) and (\ref{HH3}) for
$ - \Tr\ln (\hat{H})$. We shall see that the MA is indeed a
feature in the IR and intermediate regimes, but suppressed in the
UV, in considering the same form factors calculated in this section
in the renormalization group equations approach.

\section{Renormalization group and beta-functions}

In this section, we describe the effect of MA in the
renormalization of running parameters and show that
the beta-functions confirm the generalized version
of Appelquist and Carazzone decoupling theorem
\cite{AC}. We shall consider the calculation
of beta-functions in the mass dependent
scheme\footnote{See, for example, \cite{ramond,mano}.}
(in contrast to Minimal Substraction Scheme). From
the form factor (polarization operator), we subtract the
counterterm at the momentum $p^2 = M^2$, with $M$ being
the renormalization point.

In order to obtain the beta-function for the effective
coupling $\eta$, we apply then the operator
$$
-\lim\limits_{n\to 4} p\frac{d}{dp} =
\lim\limits_{n\to 4} \frac{4 - a^2}{4}\,a\frac{d}{da}
$$
to the form factor $k_{SS}(a)$,
where we have used $u = p^2/m^2 = 4a^2/(4-a^2)$.
Let us show the result of this calculation for the
first method as described in the previous section.
We achieve thus the 1-loop beta-function
\beq
\beta_{1\eta} = \frac{\eta^2}{(4\pi)^2}\left\{
\frac{12-2a^2}{3a^2} + \frac{4-a^2}{a^3}
\ln\left(\frac{2-a}{2+a}\right) \right\}\,,
\eeq
while in the second method we get
\beq
\beta_{1\eta}^{(1)} = \frac{\eta^2}{(4\pi)^2}\left\{
\frac{7a^2-24}{3a^2} - \frac{a^4-12a^2+32}{4a^3}
\ln\left(\frac{2-a}{2+a}\right) \right\}\,.
\eeq
The above results are quite discrepant, but
they provide the same UV limit, which is computed
directly by calculating the limit for $a\to 2$.
The result following the third and fourth methods
are identical because both form factors related to
$S_{\mu\nu}^2$-term are identical. We have then
\beq
\beta_{1\eta}^{(2,3)} = \frac{\eta^2}{(4\pi)^2}\left\{
\frac{a^2+12}{12a^2} - \frac{a^4-16}{16a^3}
\ln\left(\frac{2-a}{2+a}\right) \right\}\,.
\eeq
Despite the discrepancy between these three results,
in the UV regime we find the same beta-functions (which
coincides with the beta-function in Minimal Subtraction
Scheme). We obtain
\beq
\beta_{1\eta\, UV} = \beta_{1\eta\, UV}^{(1)} =
\beta_{1\eta\, UV}^{(2,3)} =
\frac{2}{3}\frac{\eta^2}{(4\pi)^2}\,.
\eeq

We should mention, however, that in the IR limit
(i.e., $a\to 0$),
the beta-functions describe in all cases the
Appelquist and Carazzone decoupling theorem
with different behaviors:
\beq
\beta_{1\eta\, IR} & = & \frac{1}{15}\frac{\eta^2}{(4\pi)^2}\frac{p^2}{m^2} +
{\cal O}\left(\frac{p^4}{m^4}\right)\,, \\
\beta_{1\eta\, IR}^{(1)} & = & \frac{1}{5}\frac{\eta^2}{(4\pi)^2}\frac{p^2}{m^2} +
{\cal O}\left(\frac{p^4}{m^4}\right)\,, \\
\beta_{1\eta\, IR}^{(2,3)} & = & \frac{1}{10}\frac{\eta^2}{(4\pi)^2}\frac{p^2}{m^2} +
{\cal O}\left(\frac{p^4}{m^4}\right)\,.
\eeq
In above formulas, the results are expressed in terms of the squared external
momentum $p^2$ and the fermion mass $m$. Notice that the most dominant terms
are proportional to $p^2/m^2$ which is already very small in the IR limit
($p^2 << m^2$). So, we conclude that the Appelquist and Carazzone theorem
definitely holds in the case of torsion, but the coefficients of
${\cal O}\left(\frac{p^2}{m^2}\right)$ depends on the calculational scheme.
This situation is pretty much the same as for QED \cite{bruno} and
Yukawa model \cite{dante}.

Nevertheless, one can find an intriguing feature in our results,
in particular on the expressions for the form factors
$(\na_\mu S^\mu)\left[ k_{\na S}(a) \right]\,(\na_\nu S^\nu)$
and $S^2\left[k_{S4}(a) \right]\,S^2$. As
a confirmation of known results (see, e.g., Ref. \cite{2000}),
there are no one-loop divergent terms $\sim (\na_\mu S^\mu)^2$ and
$\sim (S_\mu S^\mu)^2$; although we find here the appearance of finite
non-local corrections.

That is a very unusual feature in considering the issue of the
decoupling mechanism. For instance, there are no such beta-functions
in MS scheme
corresponding to these interactions, indeed in the UV limit
the application of $p\frac{d}{dp}$ to $k_{\na S}(a)$ and
$k_{S4}(a)$ vanishes in both cases and in all schemes.
However, in intermediate and IR regimes, the beta-functions
are non-trivial (and in fact are sensitive to calculational
scheme, or MA). Thus, in these cases the word {\it decoupling}
is not clear (because not only in the IR, the beta-functions
go to zero as ${\cal O}\left(\frac{p^2}{m^2}\right)$, but also in UV
they go to zero).

One should mention that, in Ref. \cite{2000},
it was shown that the spin-0 related term $(\na_\mu S^\mu)^2$
did not appear in the one-loop level, in a very different scenario
where fermions and massive dynamical torsion are all quantized,
but appeared in the two-loop level thanks to some specific
form of the one-loop torsion quantum effects. In the present work,
we get the result that this renormalizability and unitarity
breaking term, $(\na_\mu S^\mu)^2$, manifests already in one-loop
level.

\section{Conclusions and final remarks}

We have confirmed real inconsistency in the formula
$\Tr\ln(\hat{H}_1\,\hat{H}_2) = \Tr\ln(\hat{H}_1)
+ \Tr\ln(\hat{H}_2)$ when applied to massive fermions
coupled with an external torsion field. This feature
was presented for other cases in previous papers
\cite{bruno,dante}. The ambiguity, called non-local
Multiplicative Anomaly (MA), is supressed in calculation
of Schwinger-DeWitt coefficient $a_2$, in four dimensions,
which corresponds to local 1-loop logarithmic divergences.

The UV regime is not sensitive to MA, in contrast to
what happens in other regimes, specially in the IR. It
is typically a feature coming from the finite non-local
part of quantum corrections, as described by the expansion
of $\Tr\ln(\hat{H})$ in powers of curvature. It turns out
that the beta-functions are affected by MA in the IR and
intermediate regimes, but are universal in the UV.
We found also finite non-local contribution to
some beta-functions which are absent in UV limit,
a feature which deserves further attention.

It is interesting that this universality in the UV limit
could be derived directly from the finite non-local
1-loop effective action, in the same time as we know that
the universality of beta-functions in the UV comes from
the usual consideration of local logarithmic divergences,
which are unambiguous. Thus, in our opinion, this is
an indication that all possible ambiguity plaguing the results
comes from the MA.

%%%%%%%%%%%%%%%%%%%%%%%%%%%%%%%%%%%%%%%%%%%%%%%%%%%%%%%%%%%%%%%
\section*{Acknowledgements}

G.B.P. is grateful to CNPq and FAPEMIG
for partial support. A.E.M. is grateful to CAPES for the PhD
support program. The authors acknowledge I.L. Shapiro for
suggesting to study the MA for fermions in external torsion,
as well as reading the manuscript.

%%%%%%%%%%%%%%%%%%%%%%%%%%%%%%%%%%%%%%%%%%%%%%%%%%%%%%%%%%%
%%%%%%%%%%%%%%%%%%%%%%%%%%%%%%%%%%%%%%%%%%%%%%%%%%%%%%%%%%%
%%%%%%%%%%%%%%%%%%%%%%%%%%%%%%%%%%%%%%%%%%%%%%%%%%%%%%%%%%%

%%%%%%%%%%%%%%%%%%%%%%%%%%%%%%%%%%%%%%%%%%%%%%%%%%%%%%%%%%%

\end{document}